\documentclass[12pt]{iopart}
\usepackage[utf8]{inputenc}
\usepackage{iopams}
\usepackage{graphicx}
\usepackage{color}
\usepackage{subcaption}
\usepackage{bm}
\usepackage{hyperref}
\usepackage{float}
\expandafter\let\csname equation*\endcsname\relax
\expandafter\let\csname endequation*\endcsname\relax
\usepackage{amsmath}
\usepackage{amssymb}
\hypersetup{colorlinks=true, pdfstartview=FitV, linkcolor=blue, citecolor=black}
\usepackage[all]{hypcap}
\usepackage{multicol}
\def\no{\nonumber}

\begin{document}

\title[Controlling the surface roughness in the one-dimensional KPZ growth process]{The role of the non-linearity in controlling the surface roughness in the one-dimensional Kardar--Parisi--Zhang growth process}

\author{Priyanka$^\dagger$, Uwe C T{\"a}uber$^\dagger$ \& Michel Pleimling$^{\dagger \ddagger}$}

\address{$\dagger$ Department of Physics (MC 0435) and Center for Soft Matter and Biological Physics, Virginia Tech, Blacksburg, VA 24061, USA}
\address{$\ddagger$ Academy of Integrated Science (MC 0563), Virginia Tech, Blacksburg, VA 24061, USA}
\ead{pri2oct@vt.edu}
\vspace{10pt}

\begin{abstract}
We explore linear control of the one-dimensional non-linear Kardar--Parisi--Zhang (KPZ) equation with the goal to understand the effects the control process has on the dynamics and on the stationary state of the resulting stochastic growth kinetics.
In linear control, the intrinsic non-linearity of the system is maintained at all times.  
In our protocol, the control is applied to only a small number $n_c$ of Fourier modes.
The stationary-state roughness is obtained analytically in the small-$n_c$ regime with weak non-linear coupling wherein the controlled growth process is found to result in Edwards--Wilkinson dynamics. 
Furthermore, when the non-linear KPZ coupling is strong, we discern a regime where the controlled dynamics shows scaling in accordance to the KPZ universality class.
We perform a detailed numerical analysis to investigate the controlled dynamics subject to weak as well as strong non-linearity.
A first-order perturbation theory calculation supports the simulation results in the weak non-linear regime.
For strong non-linearity, we find a temporal crossover between KPZ and dispersive growth regimes, with the crossover time scaling with the number $n_c$ of controlled Fourier modes.
We observe that the height distribution is positively skewed, indicating that as a consequence of the linear control, the surface morphology displays fewer and smaller hills than in the uncontrolled growth process, and that the inherent size-dependent stationary-state
roughness provides an upper limit for the roughness of the controlled system.
\end{abstract}

\submitto{\JPA}
\maketitle

\section{Introduction}
Surface roughening plays an essential role in understanding the scale-invariant properties of many physical, chemical, and biological phenomena. The development of various advanced and optimized technologies requires an understanding of surface roughening phenomena, {\it e.g.}, in medicine, the fabrication of nanomaterials and biofilms for nanodevices. The ability to control the surface roughness is a critical aspect in these applications \cite{Baer13, Makeev02, LI20, Villapun20, Elsholz:2004}. One of the key challenges encountered during the control experiments is to maintain the inherent dynamical properties of the system while keeping as many stationary-state features as possible unaltered during the process to achieve a specified target. The most common reason for these difficulties is the presence of non-linearly interacting modes that are also critical for characterizing the dynamics of the growth process. Hence, identifying an appropriate control scheme and investigating its effect on the system's inherent non-linearity is important. In previous studies on interface dynamics, a wide variety of theoretical models have been used to understand the growth of the surface roughness, which can all be suitably described by stochastic partial differential equations~\cite{Barabasi95, Kuramoto78, Sivashinsky80, Kardar86, Amar:1993, Halpin-Healy:1995}.
   
The Kardar--Parisi--Zhang equation (KPZ) \cite{Kardar86}, one of the most widely used surface growth stochastic partial differential equations, has been successfully utilized to explain observations from a broad variety of experiments as well as from numerical simulations of particle-based models \cite{Takeuchi:2011, Fukai17, Halpin2015, Kriecherbauer:2010}. In one spatial dimension the KPZ growth of the interface height $h(x,t)$ with time $t$ at any given position $x$ is described by
\begin{equation}
\frac{\partial h}{\partial{t}}=\frac{\partial^2{h}}{\partial{x^2}}+\frac{g}{2}\left|\frac{\partial{h}}{\partial{x}}\right|^2+\eta(x,t)~,
\label{KPZ_eq}
\end{equation}
%g=\lambda\sqrt{\sigma/\nu^3} is needed
where $g$ denotes a scaled non-linear coupling constant, whereas $\eta(x,t)$ represents delta-correlated white noise of unit strength, i.e., $\langle\eta(x,t)\rangle=0$,
$\langle\eta(x',t')\eta(x,t)\rangle=\delta(x-x')\delta(t-t')$. Often measured is the surface roughness, which for a system of size $L$ is defined as
\begin{equation}
 W(L,t)=\sqrt{\sum_{x=-L/2}^{L/2}\langle[ h(x,t)-\langle h\rangle]^2\rangle}~,
\end{equation}
where $\langle h \rangle = \left\langle \sum_{x=-L/2}^{L/2} h(x,t) \right\rangle$ is the mean height, and the angular bracket represents averaging over stochastic realizations. Due to the self-similar properties of the growth process, the surface roughness displays the well-known Family--Vicsek scaling~\cite{Family85}, $W(L,t)\sim t^{\beta}~{\cal{W}}(t/L^{z})$, which implies that the height fluctuations grow with time as $t^{\beta}$ until they reach saturation at $t \propto L^{z}$, with the resulting surface roughness being proportional to $L^{z\beta}$ at saturation.
% the height-height correlation, $C(t,x)~=\langle h(x,t)h(0,t)\rangle$ 

The dynamical exponent $z$ and the growth exponent $\beta$ characterize the universality class of the growth process. For the KPZ equation in 1+1 dimension these exponents are exactly known, $z=3/2$ and $\beta=1/3$ \cite{Forster77}.
When setting $g=0$, the growth equation \eref{KPZ_eq} becomes linear and follows the Edwards--Wilkinson universality class \cite{Edwards82}, which can be exactly solved in any dimension with the exponents $z=2$ and $\beta=1/4$.

In this work, we use the Kardar--Parisi--Zhang equation~\cite{Kardar86} to explore a linear control scheme that saturates the roughness to a prescribed value.
We wish to shed light on the role of the non-linear coupling $g$ for reaching the preferred saturation roughness during the control process. The majority of prior research on stochastic control involved theoretical models that suppress the non-linearity either entirely or partially~\cite{Armaou00,Lou03,Lou06,Gomes15,Gomes17, Priyanka20}. Eliminating the effects of the non-linearity during the control process significantly impacts the dynamical behavior of the surface during the growth kinetics. Thus, a detailed understanding of the effects of the non-linearity during the control process on both the dynamical and steady-state properties is needed. 
In our previous work ~\cite{Priyanka20}, employing a non-linear control scheme, we demonstrated non-trivial scaling related to the KPZ universality class, which determines the time scale over which the controlled stochastic growth process is driven away from its intrinsic scale-free growth. In contrast, here we present a linear control protocol that maintains the inherent growth dynamics identical to that of the uncontrolled system, while also enforcing a desired saturation width. This kind of control approach has been previously employed for the stochastic Kuramoto--Sivashinsky equation~\cite{Lou06}, where the surface roughness control was established for just one value of the roughness saturation starting from different initial conditions. Reference~\cite{Lou06}, however, does not discuss the dynamical differences between linear and non-linear control protocols. Neither do the authors comment on the scheme's robustness with respect to modifying its input parameters, i.e., the stability of their results in reaching the prescribed roughness. 

We present in the following a detailed numerical study to elucidate the effects of control on the dynamics and the stationary-state properties for a wide range of non-linear strengths $g$ in the KPZ equation \eref{KPZ_eq}. With the help of additional analytical understanding based on first-order perturbation theory, we can determine the saturation value in some parametric regimes where a certain combination of the non-linear coupling and some control parameters is small. Our results confirm that this control scheme maintains the inherent KPZ universality class scaling features even for the controlled growth process. We believe that this result will be similarly valid for, e.g., the Kuramoto--Sivashinsky equation as it too features the KPZ non-linearity and also asymptotically falls in the KPZ universality class~\cite{Ueno:2005}.

This manuscript is organized in the following way: In section \ref{Model}, we introduce the linear control scheme with periodic control. Section \ref{Analytical} contains our perturbative calculation to obtain the eigenvalue that determines the saturation width. In section~\ref{Results}, we present our numerical results to show that the control scheme can achieve the desired width and maintain the inherent KPZ scaling properties. We conclude in section \ref{Conclusions}.

\section{Linearly controlled KPZ equation}
\label{Model}
We consider the KPZ equation~\eref{KPZ_eq} in $1+1$ dimension to study the control of the roughness of a surface during a curvature-driven stochastic growth process. As was done in the previous studies~\cite{Lou06,Priyanka20}, we perturb the uncontrolled KPZ equation with a spatially distributed linear control represented by $b_n(x)$, whence the controlled KPZ equation takes the following form:
\begin{equation}
\frac{\partial h}{\partial t}=\frac{\partial^2{h}}{\partial{x^2}}+\frac{g}{2}\left|\frac{\partial{h}}{\partial{x}}\right|^2
+ \sum\limits_{n=-n_c, n \ne 0}^{n_c} b_n(x) u_n(t) +\eta(x,t)~,
\label{controlled_KPZ_eq}
\end{equation}
where the control function $u_n(t)$ for the manipulated input $n$ is distributed over space using the actuator $b_n(x)$. We choose $b_n(x)=\exp \left(-i 2 \pi n x / L \right)$, as in our previous study (see Ref.~\cite{Priyanka20} for details); the number of actuators is $2n_c$. The controlled equation \eref{controlled_KPZ_eq} is solved with periodic boundary conditions, 
 \begin{equation}
  \frac{\partial^m h(-L/2,t)}{\partial x^m}=\frac{\partial^m h(L/2,t)}{\partial x^m} ~, \quad m=0,1,2,...
  \label{Real_CF}
 \end{equation}
In our previous work~\cite{Priyanka20}, we have shown that removing the non-linearity with different cutoff Fourier modes $n_c$ will enable the system to saturate at the desired roughness, but this significantly alters the growth dynamics. However, in order to implement a control scheme that does not change the intrinsic dynamics, it is essential to consider the effect of the non-linearity at all times. Thus, we define a linear control function as
\begin{equation}
 \sum\limits_{n=-n_c, n \ne 0}^{n_c} b_n(x) u_n(t) = - \Big[\lambda_c+\frac{\partial^2}{\partial x^2}\Big]h(x,t)~.
 \label{LControl}
\end{equation}
Here, the eigenvalue $\lambda_c$ is obtained from the the stationary roughness of the controlled KPZ equation \eref{controlled_KPZ_eq}, see Ref.~\cite{Priyanka20},
\begin{equation}
  w_d(L)=\sqrt{\frac{1}{L}\sum_{k=1}^{n_c}\frac{1}{\lambda_c}+\sum_{k=n_c+1}^{L/2}\frac{L}{(2\pi k)^2}}~,
  \label{Gauss_width}
 \end{equation}
where $w_d$ denotes the desired surface roughness and $L$ the finite system size. To start our analysis, we first perform a Fourier transform of the height function as $\bar{h}(k,t)=\sum_{x=-L/2}^{L/2}h(x,t) e^{i2\pi k x/L}$, with the total number $L$ of Fourier modes. Further, the Fourier transform of the exponential distribution of the actuators $b_n(x)$ simply becomes a Kronecker delta, $\hat{b}_{nk}=\sum_{x=-L/2}^{L/2}b_n(x) e^{i2\pi k x/L}=\delta_{n,k}$. Consequently, eq.~\eref{LControl} reduces to
\begin{equation}
\sum_{n=-n_c, n \ne 0}^{n_c}\delta_{n, k}=-\left[\lambda_c-\left(\frac{2\pi k}{L}\right)^2\right]\bar{h}(k,t)~.
\end{equation}
Using the above expression, we obtain a set of $L$ controlled coupled equations for the Fourier modes:
\begin{eqnarray}
\partial_t {\bar{h}}(k,t)& = & - \lambda_c {\bar{h}}(k,t) + Q(k,{\bar{h}})+\bar{\eta}(k,t)~, ~~ \mbox{if} ~ 0 < \left|k \right| \le n_c \label{modes_c} ~,\\
\partial_t {\bar{h}}(k,t)& = &  -\left(\frac{2\pi k}{L}\right)^2 {\bar{h}}(k,t) +Q(k,{\bar{h}}) + \bar{\eta}(k,t)~, \ \mbox{if} ~ L/2 \ge \left|k \right| > n_c~. 
\label{modes_u}
\end{eqnarray}
In the expression above, $Q(k,{\bar{h}})$ is the convolution sum of the non-linear term of the KPZ equation,
\begin{equation}
Q(k,{\bar{h}})=-\frac{g}{2}\sum_{q,q'}(q\cdot q'){\bar{h}}(q,t){\bar{h}}(q',t)\delta_{q+q',k}~,
\label{NLKPZ}
\end{equation}
and  $\bar{\eta}(k,t)$ is again delta-correlated Gaussian white noise with zero mean and variance; i.e., $\langle\bar{\eta}(k,t)\rangle=0$, $\langle\bar{\eta}(k,t)\bar{\eta}(k',t')\rangle=L\delta_{k,-k'}\delta(t-t')$. Generally, the above equations can be solved numerically. However, in the limit of weak non-linearity, an approximate analytical solution is possible. We solve the system of equations \eref{modes_c}--\eref{NLKPZ} using a pseudo-spectral method where the non-linear and noise terms are treated using a low-storage third-order Runge--Kutta scheme, and the linear diffusion term is solved using the standard Crank--Nicolson scheme. The details of the numerical method are described in Refs.~\cite{Spalart91, Priyanka20}. 

\section{Perturbative calculation}
\label{Analytical}
The numerical solution obtained with the eigenvalue $\lambda_c$ from eq.~\eref{Gauss_width} is not robust with respect to all parameters. In the following we perform a first-order regular perturbation theory analysis that allows us to study in a systematic way the effect of a weak non-linearity on the resulting surface roughness.

The presence of the non-linear term in the linear controlled KPZ equation \eref{modes_c} creates complex mixing of the controlled and uncontrolled Fourier modes. The complete separation of these coupled modes is a tedious task. However, we can use a perturbative approach to solve the problem in a weak non-linear regime. 
To start, in momentum space we add the label $c$ to the height function for the controlled modes, i.e. $\bar{h}_c(k,t) = \bar{h}(k,t)$ for $-n_c \le k < 0$ and $0 < k \le n_c$, and the label $u$ for the
uncontrolled  modes, i.e. $\bar{h}_u(k,t) = \bar{h}(k,t)$ for $k=0$, $-L/2 \le k < -n_c -1$, and $n_c+1 < k \le L/2$. In addition, we define $\bar{h}_c(k,t) = 0$ for $|k|>n_c$ and $k=0$, as well as $\bar{h}_u(k,t) = 0$ for $0<|k|\le n_c$.
%To start, in momentum space we define the height function for the modes $0$ to $|n_c|$ as $\bar{h}_c(k,t)$ and for the modes $|n_c+1|$ to $|L/2|$ as $\bar{h}_{u}(k,t)$. We also define $\bar{h}_c(k,t)=0$ for $|k|>n_c$ and $\bar{h}_u(k,t)=0$ for $0<|k|\le|n_c|$. 
To simplify, we perform the temporal Laplace transform: $\hat{h}(k,s)=\int_0^\infty \bar{h}(k,t) \exp(-s~t)~dt$.
Then the controlled dynamical equation \eref{modes_c} yields the following coupled equations in the range from $-n_c$ to $n_c$ (except $k=0$)~\cite{Frey94, Bhattacharjee07},
\begin{align}
s \hat{h}_c(k,s)&=-\lambda\hat{h}_c(k,s)-\frac{g}{2}\sum_{p=-L/2}^{L/2}\sum_{s_1=-\infty}^{\infty}p(k-p)\hat{h}_c(p,s_1)\hat{h}_c(k-p,s-s_1)\no\\
&\quad -\frac{g}{2}\sum_{p=-L/2}^{L/2}\sum_{s_1=-\infty}^{\infty}p(k-p)\hat{h}_c(p,s_1)\hat{h}_u(k-p,s-s_1)\no\\
&\quad - \frac{g}{2}\sum_{p=-L/2}^{L/2}\sum_{s_1=-\infty}^{\infty}p(k-p)\hat{h}_u(p,s_1)\hat{h}_c(k-p,s-s_1)\no\\
&\quad - \frac{g}{2}\sum_{p=-L/2}^{L/2}\sum_{s_1=-\infty}^{\infty}p(k-p)\hat{h}_u(p,s_1)\hat{h}_u(k-p,s-s_1)+\hat{\eta}(k,s)~.
\label{control_dyn0}
\end{align}
Similarly, for the uncontrolled Fourier modes with $|k| > n_c$ the dynamical equation \eref{modes_u} becomes
\begin{align}
s \hat{h}_{u}(k,s)&=-k^2\hat{h}_{u}(k,s)-\frac{g}{2}\sum_{p=-L/2}^{L/2}\sum_{s_1=-\infty}^{\infty}p(k-p)\hat{h}_{u}(p,s_1)\hat{h}_u(k-p,s-s_1)\no\\
&\quad -\frac{g}{2}\sum_{p=-L/2}^{L/2}\sum_{s_1=-\infty}^{\infty}p(k-p)\hat{h}_{c}(p,s_1)\hat{h}_{u}(k-p,s-s_1)\no\\
&\quad -\frac{g}{2}\sum_{p=-L/2}^{L/2}\sum_{s_1=-\infty}^{\infty}p(k-p)\hat{h}_u(p,s_1)\hat{h}_c(k-p,s-s_1)\no\\
&\quad -\frac{g}{2}\sum_{p=-L/2}^{L/2}\sum_{s_1=-\infty}^{\infty}p(k-p)\hat{h}_{c}(p,s_1)\hat{h}_c(k-p,s-s_1)+\hat{\eta}(k,s)~.
\label{uncontrol_dyn0}
\end{align}
All the sums above are convolutions of the form given in eq.~\eref{NLKPZ}, and $\hat{\eta}(k,s)$ is the Laplace transform of $\bar{\eta}(k,t)$ with zero mean and variance
\begin{equation}
 \langle\hat{\eta}(k,s)\hat{\eta}(k',s')\rangle=2\pi \frac{\delta_{k,-k'}}{s+s'}~.
\end{equation}
At zeroth order in the non-linearity, the height function is
\begin{align}
 \hat{h}_c(k,s)&=\frac{\hat{\eta}(k,s)}{s+\lambda_c}~, ~~~\mbox{if} ~ 0 < \left| k \right| \le n_c ~,\\
 \hat{h}_u(k,s)&=\frac{\hat{\eta}(k,s)}{s+k^2},~ ~~~\mbox{if} ~ L/2 \ge \left| k \right| >  n_c ~.
 \label{Linear_h}
\end{align}
We note that due to the control, the solution for the height function in the control regime is not the same as in the uncontrolled case, and hence the spatial symmetry is broken, i.e.,
$\hat{h}_{c}(p,s_1)\hat{h}_{u}(k-p,s-s_1)\ne\hat{h}_{u}(p,s_1)\hat{h}_{c}(k-p,s-s_1)$.
The quantity of interest is the surface roughness $W(L,t)=\sqrt{\sigma(L,t)}$. The Laplace transform $\bar{\sigma}(L,s,s_1)$ of $\sigma(L,t,t_1)$ is given as
\begin{eqnarray}
 \bar{\sigma}(L,s,s_1)&=&\frac{1}{L}\sum_{k=1}^{n_c}\sum_{k_1=1}^{n_c}\langle \hat{h}(k,s) \hat{h}(k_1,s_1)\rangle\delta_{k,-k_1}\no\\
 &=&\frac{1}{L}\sum_{k=1}^{n_c}\sum_{k_1=1}^{n_c}\langle \hat{h}_c(k,s) \hat{h}_c(k_1,s_1)\rangle\delta_{k,-k_1}\no\\
 &&+\frac{1}{L}\sum_{k=n_c+1}^{L/2}\sum_{k_1=n_c+1}^{L/2}\langle \hat{h}_u(k,s) \hat{h}_u(k_1,s_1)\rangle\delta_{k,-k_1}~.
 \label{width0}
\end{eqnarray}
Arriving at a closed solution of the above equations is very tedious. For this reason, we focus only on the weak non-linear coupling regime and restrict our perturbative analysis to ${\mathcal{O}}(g^2)$ terms. This approximation nevertheless can explain an observed shift of the saturation width away from the desired value in the parameter regime of interest. 

Inserting the linear solution from eq.~\eref{Linear_h} into eq.~\eref{width0}, we arrive at the Laplace transform of the roughness up to ${\mathcal{O}(g^2)}$  comprising all time-dependent terms given by eqs.~\eref{zero_ss}--\eref{g2_UU} in \ref{appendix}. We perform all possible contractions of the noise terms and the contour integration of surviving terms to arrive at an expression for the time-dependent roughness (see \ref{appendix}). In the limit of $t\rightarrow \infty$, the square of the roughness $W(L)$ is given by 
\begin{eqnarray}
\hspace{-2.25cm}
 W^2(L,t\rightarrow\infty)&=&w_d^2(L) \no\\
 &=& \frac{2}{L}\left[\sum_{k=1}^{n_c}\frac{1}{2\lambda}+\sum_{k=n_c+1}^{L/2}\frac{1}{2 k^2}\right]-\frac{4 g^2\pi^2}{L^4}\sum_{k=1}^{n_c}\sum_{p=1}^{nc}\frac{(2\pi/L)^4 p^2 (k - p)^2}{ 3 \lambda ^4}
 \no\\&&  % Cacc
- \frac{4g^2 \pi^2}{L^4}\sum_{k=1}^{n_c}\sum_{p=n_c+1}^{L/2}\frac{\Theta(|k-p|-n_c)}{\lambda((k-p)^2+p^2)(2\pi/L)^2+\lambda^2}\no\\&& %Cauu
- \frac{4g^2 \pi^2}{L^4}\sum_{k=1}^{n_c}\sum_{p=n_c+1}^{L/2}\frac{(2\pi(k-p)/L)^2~\Theta(n_c-|k-p|)}{\lambda^2~((2\pi p/L)^2)+2\lambda)}\no\\&& %Cauc
- \frac{4 g^2 \pi^2}{L^4}\sum_{k=n_c+1}^{L/2}\sum_{p=1}^{n_c}\frac{p^2~\Theta(|k-p|-n_c)}{\lambda k^2~\left[(2\pi(k-p)/L)^2+(2\pi(k)/L)^2+\lambda\right]}\no\\&& %Uacu
- \frac{4 g^2 \pi^2}{L^4}\sum_{k=n_c+1}^{L/2}\sum_{p=n_c+1}^{L/2}\frac{(k-p)^2~\Theta(n_c-|k-p|)}{\lambda k^2~\left[(2\pi(p)/L)^2+(2\pi k/L)^2+\lambda\right]}\no\\
 && %Uauc
- \frac{4 g^2 \pi^2}{L^4}\sum_{k=n_c+1}^{L/2}\sum_{p=1}^{n_c}\frac{(2\pi p/L)^2(k-p)^2~\Theta(n_c-|k-p|)}{\lambda k^2~\left[(2\pi k/L)^2+2\lambda\right]}\no\\&& %Uacc
- \frac{ 4g^2 \pi^2}{L^4}\sum_{k=n_c+1}^{L/2}\sum_{p=n_c+1}^{L/2}\frac{\Theta(n_c-|k-p|)}{2 (2\pi/L)^4k^2~[(k-p)^2+ k p]}~, %Uacc
\label{widthformula}
\end{eqnarray}
where $\lambda$ indicates the roots obtained by setting the right-hand side of the above equation equal to the square $w_d^2(L)$ of the desired saturation roughness. We then compare these eigenvalues with the eigenvalue $\lambda_c$ evaluated from the zeroth-order, linearly approximated roughness given by eq.~\eref{Gauss_width}.
 
Analyzing all ${\mathcal{O}}({g^2})$ terms in eq.~\eref{widthformula}, we find that the contribution from the second term on the right-hand side of the equation is most important. The contributions due to the other terms are minimal and can be neglected. Thus, for weak non-linear coupling $g\le4$ with appropriate $n_c$, the square of the stationary-state saturation roughness for the controlled KPZ equations \eref{modes_c} and \eref{modes_u} can be approximated by
\begin{eqnarray}
\hspace{-1cm} w_d^2(L)&\approx& \frac{2}{L}\left[\sum_{k=1}^{n_c}\frac{1}{2\lambda}+\sum_{k=n_c+1}^{L/2}\frac{1}{2 k^2}\right]-\left(\frac{ 2\pi g}{L^2}\right)^2\sum_{p,k=1}^{n_c}\frac{ (2\pi/L)^4p^2 (k - p)^2}{ 3 \lambda^4}~.
\label{second-order_sol}
\end{eqnarray}

\begin{figure}[!ht]
\centering \includegraphics[width=\textwidth]{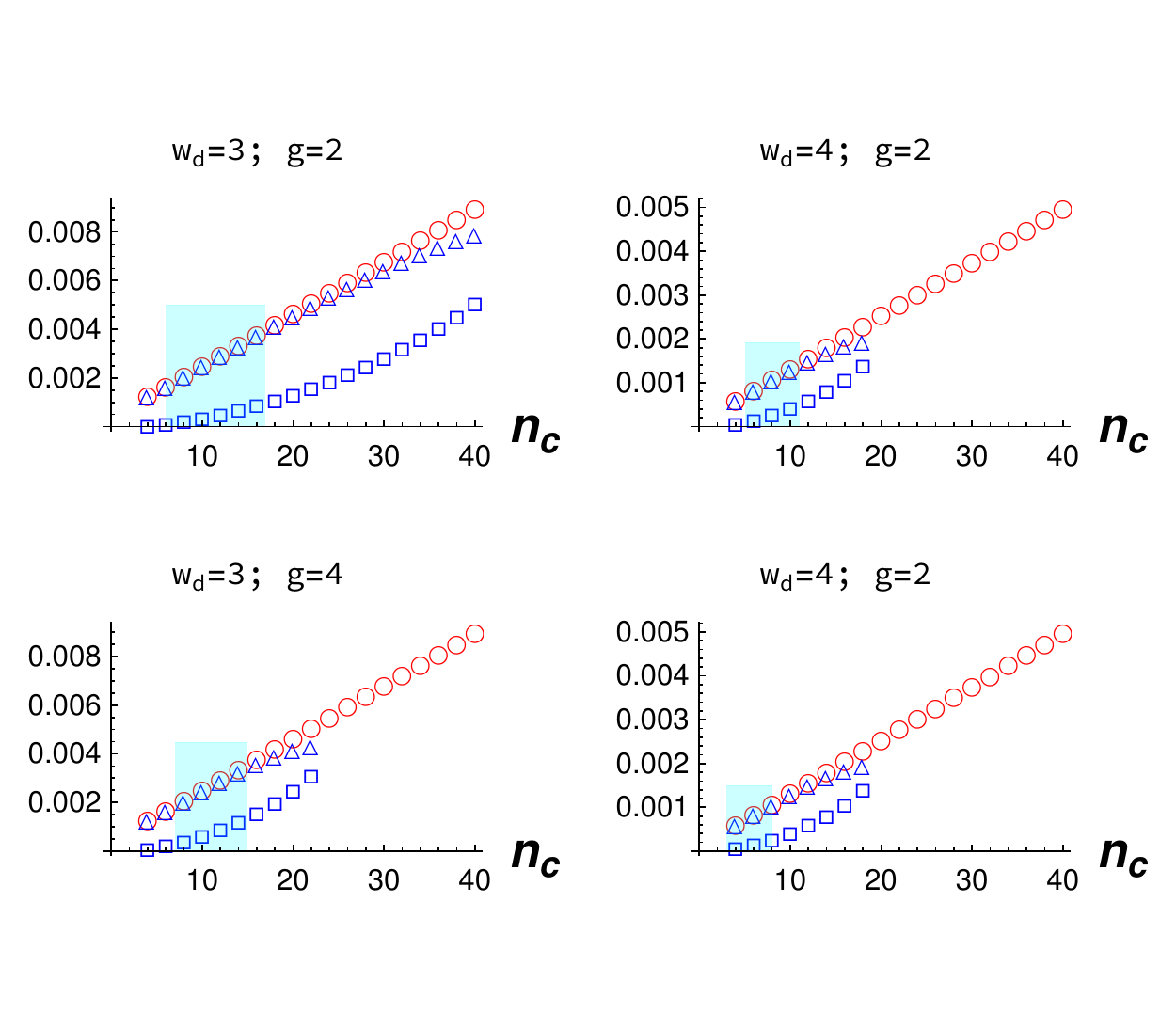}
% EigenVal.pdf: 0x0 px, 300dpi, 0.00x0.00 cm, bb=
\caption{Comparison of the roots $\lambda_c (\circ)$ (linear approximation) and $\lambda_1$ ($\vartriangle$), $\lambda_2$ ($\square$) (second-order approximation) obtained from eqs.~\eref{Gauss_width} and \eref{second-order_sol}, respectively, for different numbers $2n_c$ of controlled modes; two different desired steady-state roughness values $w_d=3,4$ are specified; upper panels: $g=2$, lower panels: $g=4$. The shaded region represents the parameter region for which the system can be controlled such that the final surface width is within $6.5 \%$ of the desired width $w_d$; the lattice size used here is $L=512$.}
\label{fig:sol_lambda}
\end{figure}

For any desired roughness $w_d$, the above equation yields four roots for $\lambda$ of which two are real and two are imaginary solutions in the small $g$ and $n_c$ parametric regime. Among the two real solutions ($\lambda_1,\lambda_2$), $\lambda_1$ has approximately the same value as $\lambda_c$ obtained from the linear solution given by eq.~\eref{Gauss_width} (see Fig.~\ref{fig:sol_lambda}). However, for large $n_c$ and sufficiently large $g$ values, all four solutions of eq.~\eref{second-order_sol} become imaginary. At this stage, we reach a point where the ${\mathcal{O}}(g^2)$ approximation is no longer adequate to determine the saturation width of the growth process, and hence the inclusion of higher-order terms would be required in the analysis. Nevertheless, our calculation provides a regime where eq.~\eref{second-order_sol} can predict the saturation width in our numerical simulation of the controlled stochastic growth process. 
  
Figure~\ref{fig:sol_lambda} presents the solution for two different values of the desired saturation roughness with non-linear couplings $g=2$ and $4$. It is clearly seen in the figure that we have a range of values for $n_c$ (the shaded cyan region), for which the linear approximation provides a reasonably good match where the deviation of the numerically obtained saturation width from the desired saturation width is less than $6.5 \%$. Beyond the shaded region, the saturation width can be predicted by eq.~\eref{second-order_sol} for a range of $n_c$ while the values of $\lambda_1$ and $\lambda_c$ are close. We would also like to mention here that in this calculation the actual perturbation parameter is not simply $g$ but instead is given by a complicated combination of $n_c$ and $g$ whose exact form is difficult to obtain. Depending on $g$, it is apparent from Fig.~\ref{fig:sol_lambda} that we may categorize the controlled KPZ equation with respect to its dynamical behavior as belonging to either of the following three regimes:
  \begin{itemize}
   \item Weak regime ($g\le2$): The uncontrolled KPZ growth is approximated by the linear Edwards--Wilkinson equation for any finite system and eq.~\eref{second-order_sol} is valid for a large range of the number $n_c$ of controlled modes.
   \item Intermediate regime ($2<g\le4$): The uncontrolled KPZ growth follows the dynamical scaling of the KPZ universality class, and the range where the ${\mathcal O}(g^2)$ approximation works is drastically diminished.
   \item  Strong regime ($g>4$): Here, one cannot predict the saturation width from eq.~\eref{second-order_sol}.
  \end{itemize}
  Our analysis also confirms that to saturate the roughness below the inherent system size-dependent saturation value of $\sqrt{L/24}$, a prior knowledge of the solution at all orders is not needed. However, in order to saturate above the limit of $\sqrt{L/24}$, a calculation including higher-order terms would need to be performed.   

\begin{figure*}[!ht]
\begin{subfigure}[b]{0.5\textwidth} 
\includegraphics[width=\textwidth]{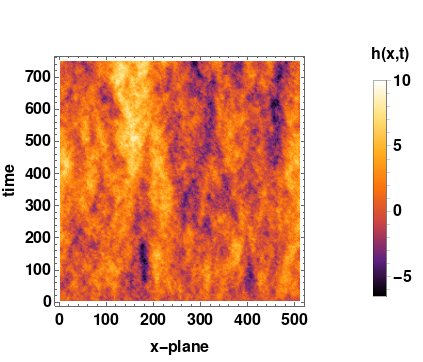}
% unHeightg1.pdf: 0x0 px, 300dpi, 0.00x0.00 cm, bb=
\subcaption{Uncontrolled case for $g=1$}
\label{g1_uh}
\end{subfigure}%
\begin{subfigure}[b]{0.5\textwidth} 
 \includegraphics[width=\textwidth]{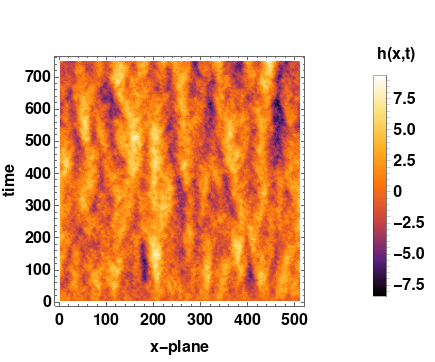}
% HeightS9g1.pdf: 0x0 px, 300dpi, 0.00x0.00 cm, bb=
\subcaption{ Controlled case for $g=1$}
\label{g1_ch}
\end{subfigure}%
\vfill
\begin{subfigure}[b]{0.5\textwidth} 
\includegraphics[width=\textwidth]{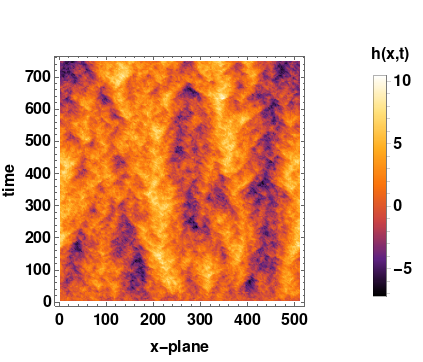}
% unHeightg1.pdf: 0x0 px, 300dpi, 0.00x0.00 cm, bb=
\subcaption{ Uncontrolled case for $g=4$}
\label{g4_uh}
\end{subfigure}%
\begin{subfigure}[b]{0.5\textwidth} 
 \includegraphics[width=\textwidth]{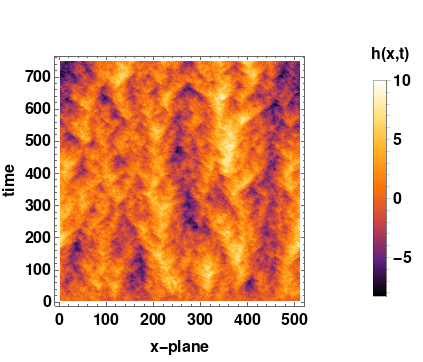}
% HeightS9g1.pdf: 0x0 px, 300dpi, 0.00x0.00 cm, bb=
\subcaption{ Controlled case for $g=4$}
\label{g4_ch}
\end{subfigure}
\caption{The plots show the surface morphology evolving in time for a single stochastic run for both controlled and uncontrolled growth processes. Each morphology starts with the same initial condition and an identical seed is used to generate the stochastic noise that determines the subsequent kinetics. The desired saturation roughness of the controlled KPZ equation is set to $w_d=3$, and $2 n_c=32$ Fourier modes are controlled.}
\label{surface_morp}
\end{figure*}
\section{Numerical results}
\label{Results}
In our previous study~\cite{Priyanka20}, we were able, using non-linear feedback control, to control the surface roughness for a broad range of saturation roughness values at the expense of severe dynamical changes which occurred at early times of the growth process. We demonstrated that the deviation time, i.e., the time period after which the controlled growth kinetics deviates from the inherent dynamics, changes algebraically with the number of controlled actuators $n_c$, with a characteristic power-law exponent that involvies the KPZ scaling exponents. Building on the understanding obtained from our earlier work with non-linear control~\cite{Priyanka20} and from the regular perturbation series analysis from Section~\ref{Analytical}, we present in this Section our numerical results obtained with linear control.

We start with a comparative study of the surface morphology obtained from linear control, eqs.~\eref{modes_c} and \eref{modes_u}, with that of the uncontrolled growth process \eref{KPZ_eq} for different values of the dimensionless non-linear strength $g$, see Fig.~\ref{surface_morp}. The morphologies shown in Fig.~\ref{surface_morp} are obtained from single runs for $L=512$ starting from an initially flat surface. The same seed value is used in all cases to generate the stochastic noise realizations. For the KPZ growth process, the uncontrolled surface looks rougher with multiple hills and valleys, irrespective of the non-linear strength, as shown in Figs.~\ref{g1_uh} and \ref{g4_uh}. Although the stochastic noise history is kept identical, we observe distinct changes in the surface morphology due to the implementation of control. Interestingly, the differences are more prominent for $g=1$ (weak non-linear strength) than for $g=4$ (see Figs.~\ref{g1_ch} and \ref{g4_ch}). In both controlled cases, we integrate the controlled KPZ equation with $n_c=16$ and set the target saturation width to $w_d=3$.

 \begin{figure*}[!ht]
 \centering
\includegraphics[width=0.95\textwidth]{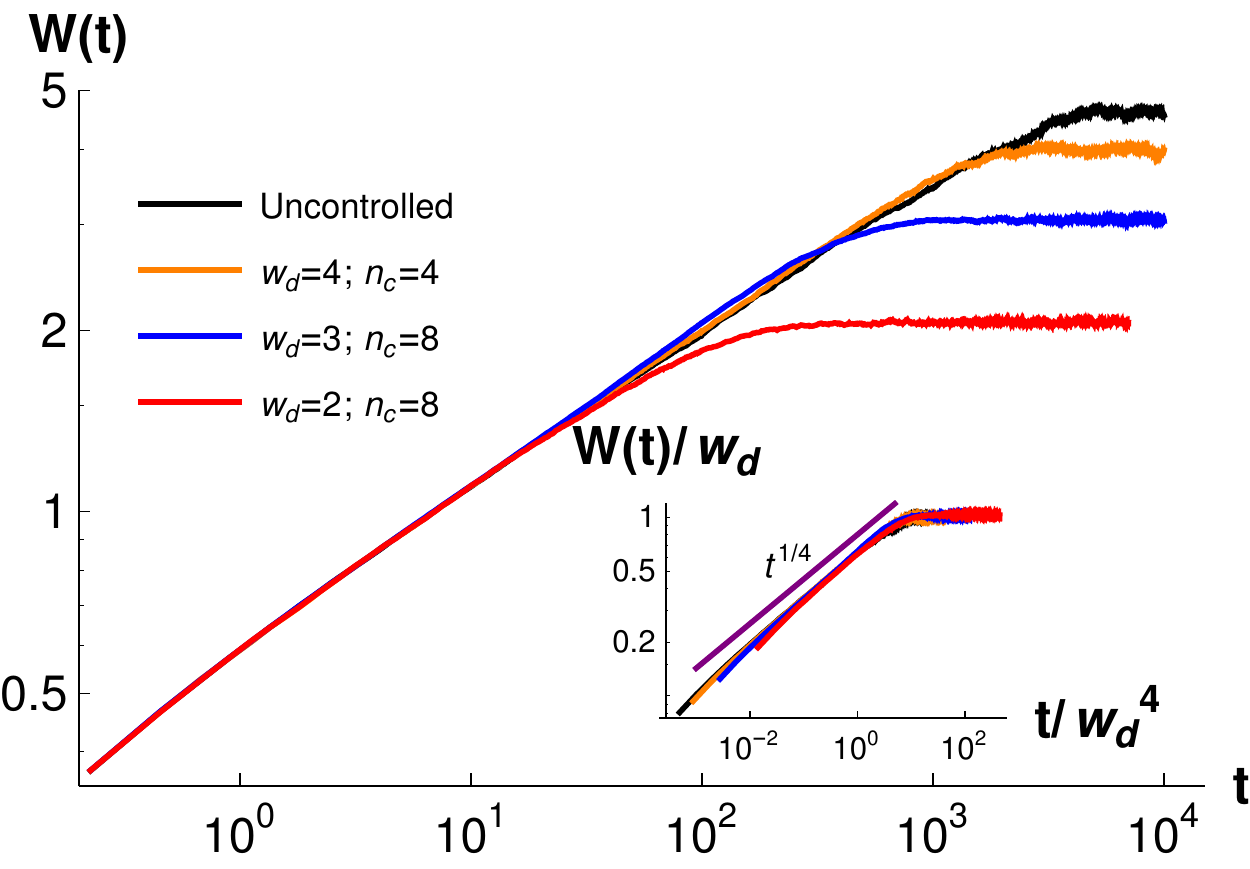}
% width_smallg.pdf: 0x0 px, 300dpi, 0.00x0.00 cm, bb=
\caption{The plot shows the controlled KPZ growth process of the surface width with weak non-linear coupling $g=1$. The main figure shows the growth of the surface roughness for various desired saturation widths. We present the scaling collapse with the desired saturation width with the EW dynamical exponent $z=2$ in the inset. The system parameters are $L=512$, $g=1$, 
where the time increment for an integration step is $\delta t = 0.0075$. The data are averaged over $3000$ different stochastic noise realizations.}
\label{g1_width}
\end{figure*}
\begin{figure*}
\centering
 \includegraphics[width=0.85\linewidth]{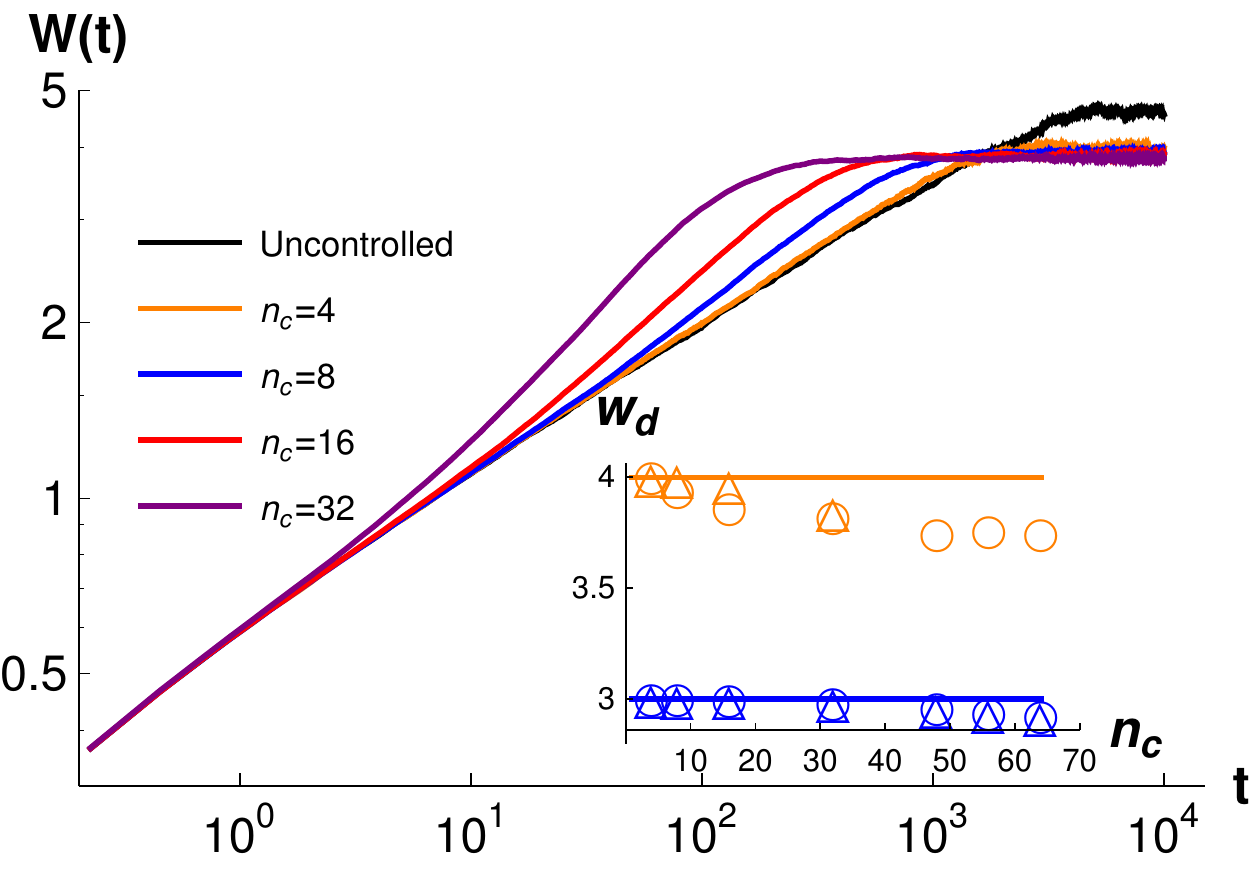}
% g1width.pdf: 0x0 px, 300dpi, 0.00x0.00 cm, bb=
\caption{The plot shows the controlled KPZ growth process of the surface width with weak non-linear coupling $g=1$. The main figure presents the change in the controlled roughness growth for various numbers $2 n_c$ of controlled modes and the common desired saturation width $w_d=4$. The inset compares the saturation widths obtained from numerical integration ($\vartriangle$) with the approximation \eref{second-order_sol} ($\circ$). Two different target widths $w_d=3$ and $w_d=4$ are considered. The system parameters are $L=512$, $g=1$, and $\delta t = 0.0075$. The data are averaged over $3000$ different noise realizations.}
\label{g1_n_cwidth}
\end{figure*}

\paragraph{Weak regime, $g\le2$:} 
In the thermodynamic limit, the solution of the KPZ equation for small $g$ converges only slowly to the non-Gaussian KPZ fixed point with its associated universal scaling exponents \cite{Ivan2012}. On the other hand, for a finite system in finite times and small $g$, the dynamics remains close to the EW fixed point. Hence for weak non-linearity the uncontrolled KPZ growth process~\eref{KPZ_eq} displays Family--Vicsek scaling for the fluctuations with EW scaling exponents. Interestingly, the numerical integration of the controlled growth process~\eref{controlled_KPZ_eq} also shows a growth of the time-dependent roughness with exponent $1/4$, as demonstrated in Fig.~\ref{g1_width} for $g=1$. Furthermore, the Family--Vicsek scaling of the controlled dynamics for different desired saturation widths confirms the presence of the same underlying EW dynamics for the controlled growth, as seen in the inset of Fig.~\ref{g1_width}. We also find that the measured dynamical exponent $z$ and roughness exponent $\alpha=z\beta$ belong to the EW universality class. The inset of Fig.~\ref{g1_width} confirms the exponents with a nice scaling collapse of the time-dependent roughness with the different saturation values $w_d$. The best collapse yields a dynamical exponent $z=2$ and roughness exponent $\alpha=1/2$. In the data collapse graph, we have also included the data points from the uncontrolled KPZ system (shown by black lines in Fig.~\ref{g1_width}) where we used $w_d=\sqrt{L/24}$. 

However, it should be noted that the EW dynamical evolution in controlled growth is limited by the number of actuators: Either a large number of controlled wavenumbers or a desired saturation roughness greater than the inherent system-size saturation width $w_d>\sqrt{L/24}$ drives the system away from its inherent dynamics. Figure~\ref{g1_n_cwidth} clearly shows that the deviation in the dynamical growth from the intrinsic behavior increases with $n_c$ for $w_d=4$. This deviation is attributed to the dispersive nature of the higher-order terms in $g$, present in the controlled equation \eref{modes_c}. However, for some range of $n_c$, the controlled dynamics is almost similar to the EW dynamics~\eref{KPZ_eq}, as shown in Fig.~\ref{g1_width}. All the results discussed in Figs.~\ref{g1_width} and \ref{g1_n_cwidth} are obtained via the controlled growth process for a fixed lattice size of $L=512$. The results are robust for systems larger than $L=128$ and remain consistent with the appropriate choice of the number of controlled modes. We further note that this scaling fails when the desired saturation width is larger than the system size saturation width $\sqrt{L/24}$, as beyond the intrinsic saturation value the controlled dynamics is fully determined by eq.~\eref{modes_c}. 
 
%%%%%%%%%%%%%%%%%%%%g=4  n_c deviation fig----------------
\begin{figure*}
    \centering
     \includegraphics[width=0.85\linewidth]{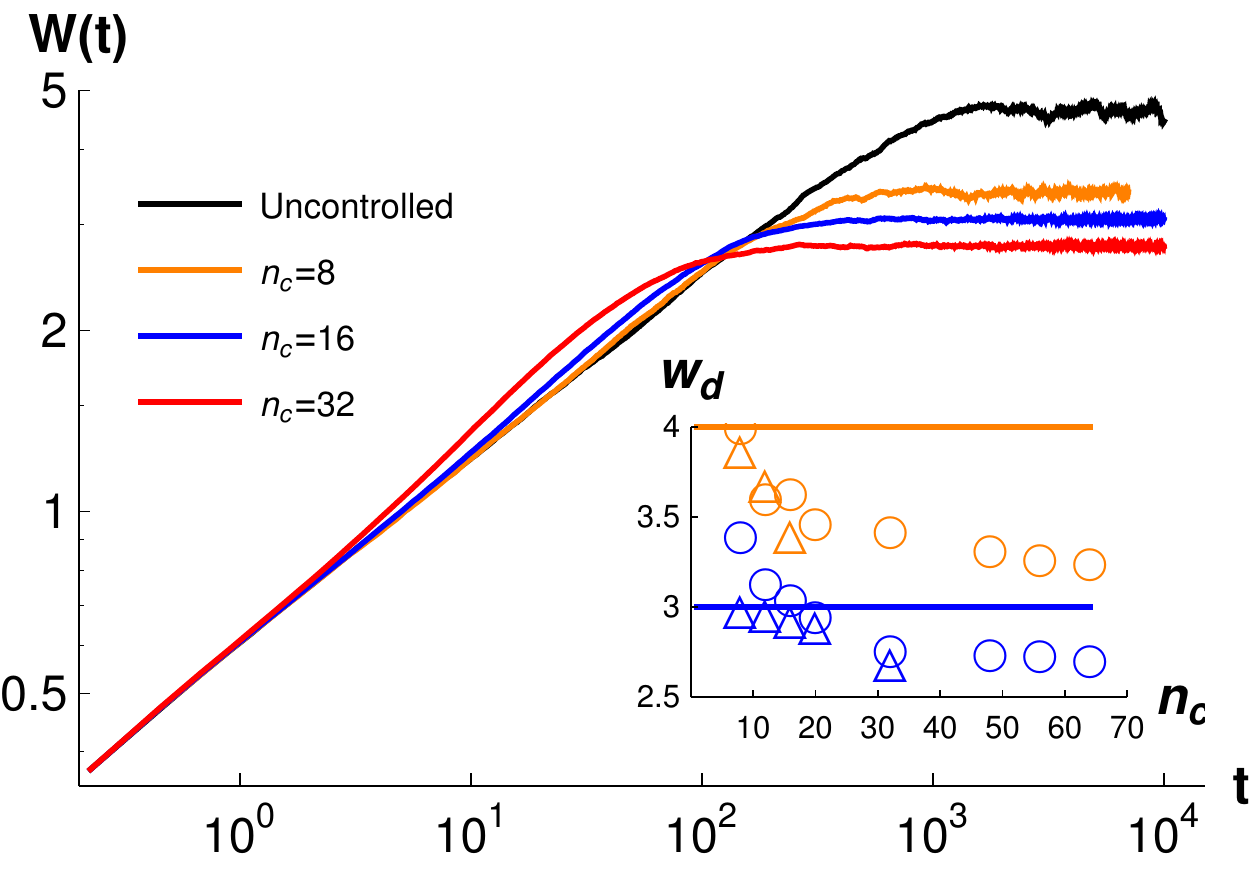}
% g4wd.pdf: 0x0 px, 300dpi, 0.00x0.00 cm, bb=
\caption{The plot displays the controlled KPZ growth process for the surface width in the intermediate non-linear coupling regime with $g=4$. The main figure presents the change in the controlled roughness growth for various numbers $n_c$ of controlled modes and the common desired saturation width $w_d=4$. 
The inset compares saturation widths obtained from numerical integration ($\vartriangle$) with the approximation (\ref{second-order_sol}) ($\circ$). Two different target widths $w_d=3$ and $w_d=4$ are considered.
The system parameters are $L=512$, $g=4$, and $\delta t = 0.0075$. The data are averaged over $3000$ different noise realizations.}
\label{fig:g4_n_cwidth}
\end{figure*}
%%%%%%%%%%%%%%%figure g=4 width%%%%%%%%%%%%%%%%%%%%%
\begin{figure*}[!ht]
\centering
 \includegraphics[width=0.85\linewidth]{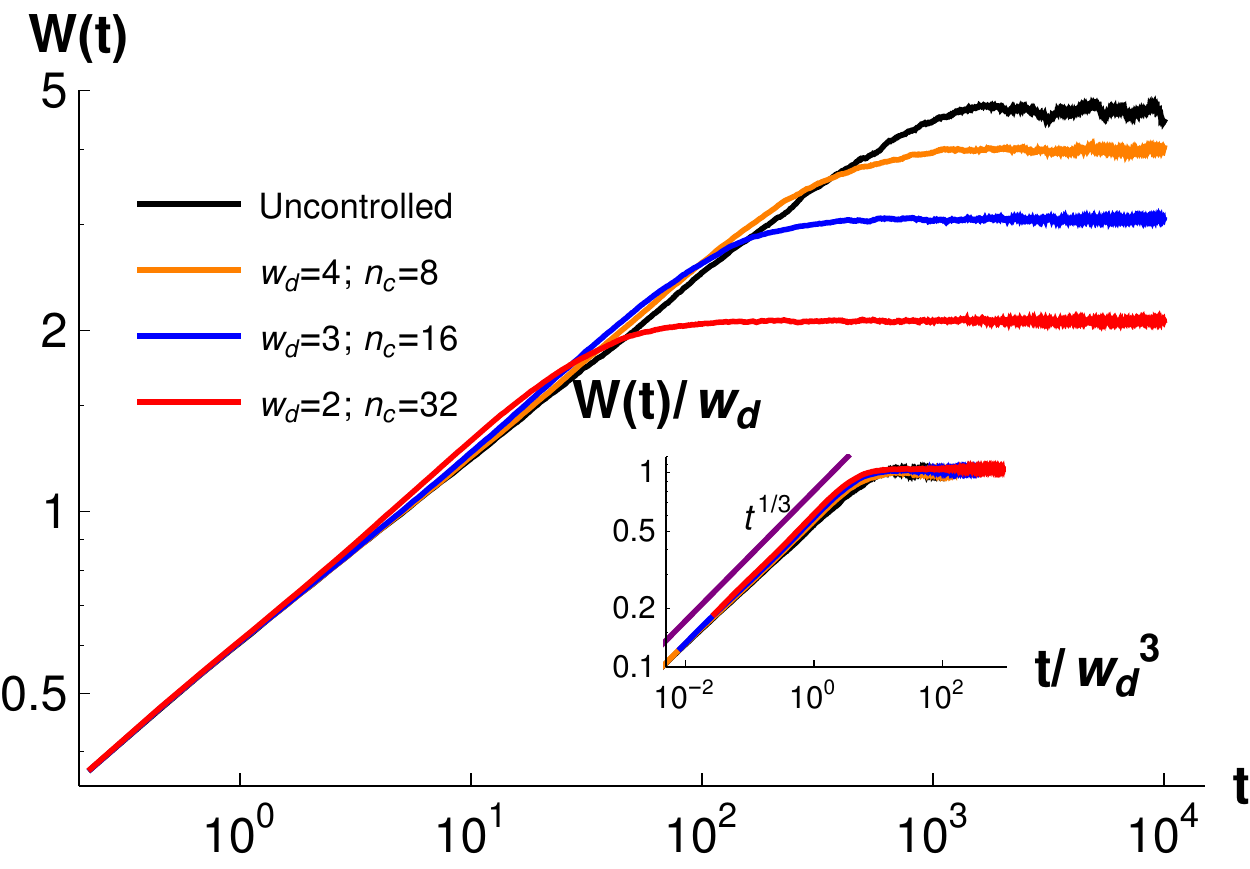}
% g4width.pdf: 0x0 px, 300dpi, 0.00x0.00 cm, bb=
\caption{The plot displays the controlled KPZ growth process for the surface width in the intermediate non-linear coupling regime with $g=4$. The main figure shows the growth of the surface roughness for various desired saturation widths. The inset presents a scaling collapse of the time-dependent roughness with the desired saturation width that involves the dynamical exponent $z=3/2$. This confirms that in this regime the controlled growth still belongs to the KPZ universality class. The system parameters are $L=512$, $g=4$, and $\delta t = 0.0075$. The data are averaged over $3000$ different noise realizations.}
\label{fig:g4_width}
\end{figure*}%

\paragraph{Intermediate regime $2<g\le4$:}
For larger non-linear couplings, we do not expect our perturbative calculation to work. Yet for intermediate strengths $g\leq4$, the perturbative approach turns out to be still acceptable in predicting the saturation values correctly in a particular range of $n_c$. As demonstrated in Fig.~\ref{fig:sol_lambda}, the shaded region for the desired roughness for $g=4$ is reduced in size. Beyond this region, the saturation value obtained from the numerics deviates drastically from the analytical calculation, as shown in the inset of Fig.~\ref{fig:g4_n_cwidth}. Moreover, in the inset of Fig.~\ref{fig:g4_n_cwidth}, we also notice that the numerical solution seems to converge to a fixed saturation width for larger $n_c$, an analytical confirmation of which would require an analysis of the dynamical equations to higher orders. Thus, we have to completely rely on numerics here. Further, our numerical integration for large non-linear strengths shows that it is not possible to saturate the surface roughness above its uncontrolled system-size dependent saturation value $\sqrt{L/24}$. The reason behind this is not apparent to us and would require more exploration as well. However, similar to the weakly non-linear case, we can extract the range of $n_c$ in the intermediate regime for which the system is controllable and reaches the desired width, as shown in Fig.~\ref{fig:g4_width}.

For small $n_c$, we find that the dynamical exponent extracted from the scaling collapse of the controlled dynamics of the growth process is compatible with the KPZ universality class.  
The inset of Fig.~\ref{fig:g4_width} shows the data collapse of the time-dependent roughness for $g=4$, with the dynamical exponent $z=3/2$, and roughness exponent $\alpha=1/2$. For this data collapse graph, we have also included the data for the uncontrolled KPZ system (shown by black lines in Fig.~\ref{fig:g4_width}), where we used $w_d=\sqrt{L/24}$. For small $n_c$, these results are robust. The discrepancy with the perturbative calculation increases with increasing $n_c$ as the perturbation approximation fails in that limit. 
%Similar deviations from the inherent dynamics can be observed in Fig.~\ref{fig:g4_n_cwidth}.

\begin{figure*}[!ht]
%\centering
    \begin{subfigure}[b]{0.5\textwidth} 
 \includegraphics[width=\textwidth]{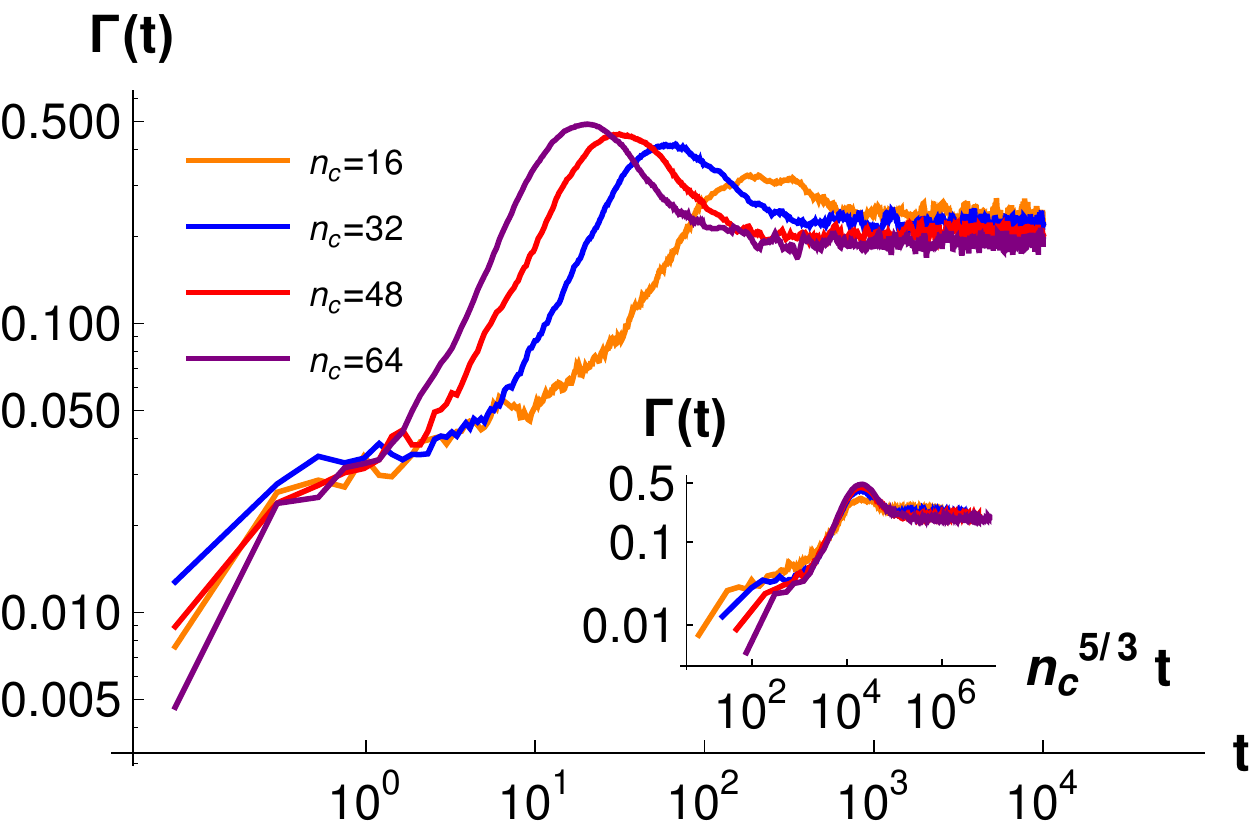}
% Skewg1.pdf: 0x0 px, 300dpi, 0.00x0.00 cm, bb=
\caption{g=1}
\label{fig:g1_skewness}
    \end{subfigure}%
     \begin{subfigure}[b]{0.5\textwidth} 
\includegraphics[width=1\textwidth]{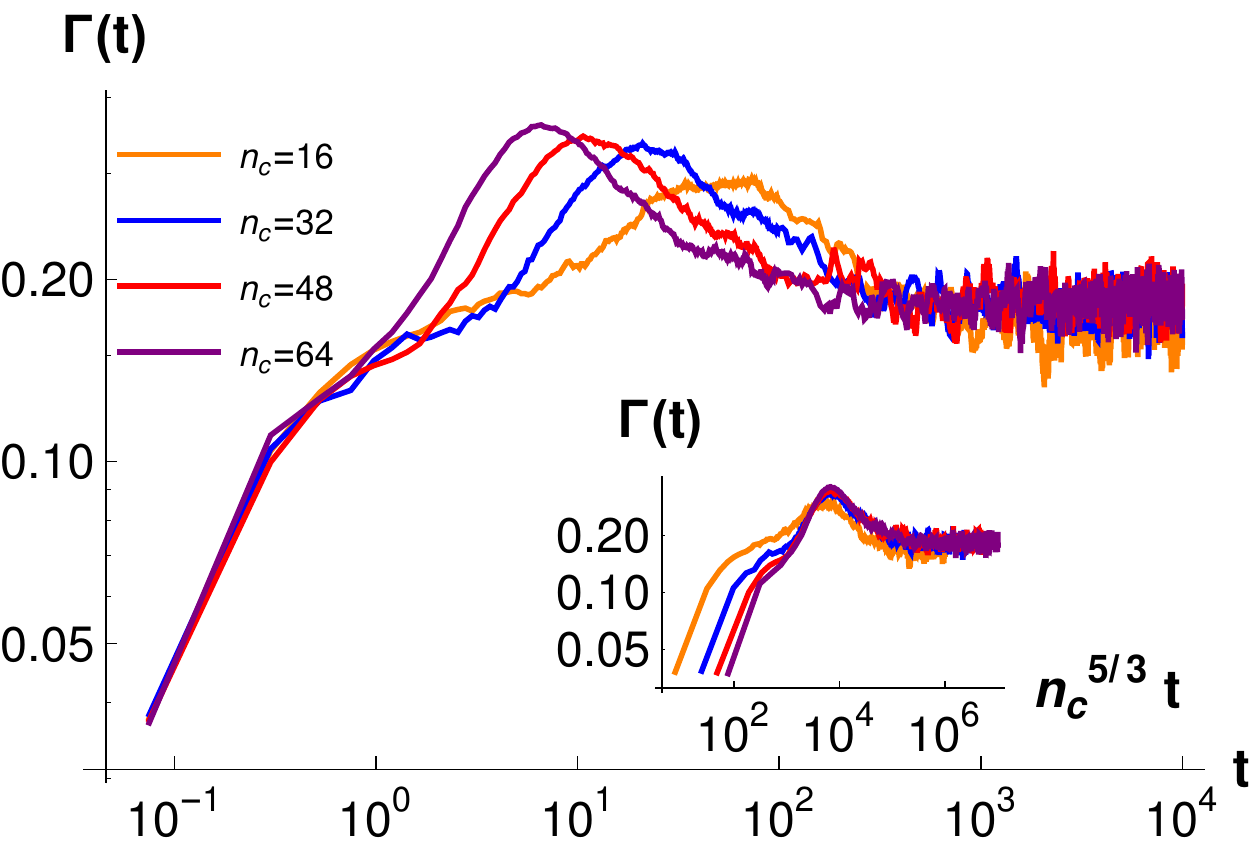}
% Skewg4.pdf: 0x0 px, 300dpi, 0.00x0.00 cm, bb=
\caption{g=4}
     \label{fig:g4_Skewness}
    \end{subfigure}
\caption{The plot shows the time-dependent skewness \eref{skewness} for the controlled KPZ growth process with the desired surface width $w_d=4$. 
The main parts of panels (a) and (b) display the change in skewness with $n_c$ for $g=1$ and $g=4$, respectively. In the insets, the scaling collapse of the skewness peak for different $n_c$ is shown with exponent $5/3$ that belongs to the KPZ universality class. The system parameters are $L=512$ and $\delta t = 0.0075$. The data are averaged over $3000$ different noise realizations.}
\label{fig:skewness}
\end{figure*}
The controlled surface growth also impacts the nature of the underlying height distribution. We study this by measuring the time-dependent skewness of the growth process, defined as
\begin{equation}
 \Gamma(t)= \left< \left( \frac{h - \langle h \rangle}{W} \right)^3 \right>,
 \label{skewness}
\end{equation}
where the angular bracket denotes averaging over all lattice sites as well as over different stochastic realizations. It is known that the height distribution of the uncontrolled KPZ growth is symmetric, and hence the skewness is zero. The skewness provides a measure for the asymmetry added to the height distribution. In our earlier study \cite{Priyanka20}, we found that non-linear feedback control skewed the distribution more towards higher $h$ values, reflecting that the surface morphology features more valleys than hills. 

In contrast, in the case of linear control presented here, the height distribution is characterized by positive skewness, i.e., dominated by lower height values, as shown in Fig.~\ref{fig:skewness}. This result is very different from the earlier studied case of non-linear control \cite{Priyanka20} as even the magnitude of the skewness turns out smaller under the linear control scheme. The positive skewness with smaller magnitude suggests that the surface morphology for linear control is more prone to have smaller hills and is concentrated closer to the mean value, as shown in Figs.~\ref{surface_morp} and \ref{fig:skewness}.
Furthermore, for large $n_c$, the skewness displays a hump-like profile during the intermediate time when the dynamics starts deviating from its inherent dynamics. This peak characterizes the time when the effect of the controlled modes on the growth dynamics kicks in. The skewness peak scales with the number of actuators $n_c$ with exponent $5/3$, the same value we observed in our previous study of non-linear control \cite{Priyanka20}. The inset of Fig.~\ref{fig:skewness} shows the collapse of the peaks for various $n_c$ and reveals that this exponent is independent of the strength of the non-linear coupling $g$. This confirms that the stochastic growth kinetics is governed by the non-linear KPZ fixed point. Our skewness analysis also suggests that we are likely to observe the inherent dynamics for the controlled growth only for those values of $g$ and $n_c$ for which the skewness peak is not prominent.

\begin{figure}[!ht]
\centering
   \includegraphics[width=0.75\linewidth]{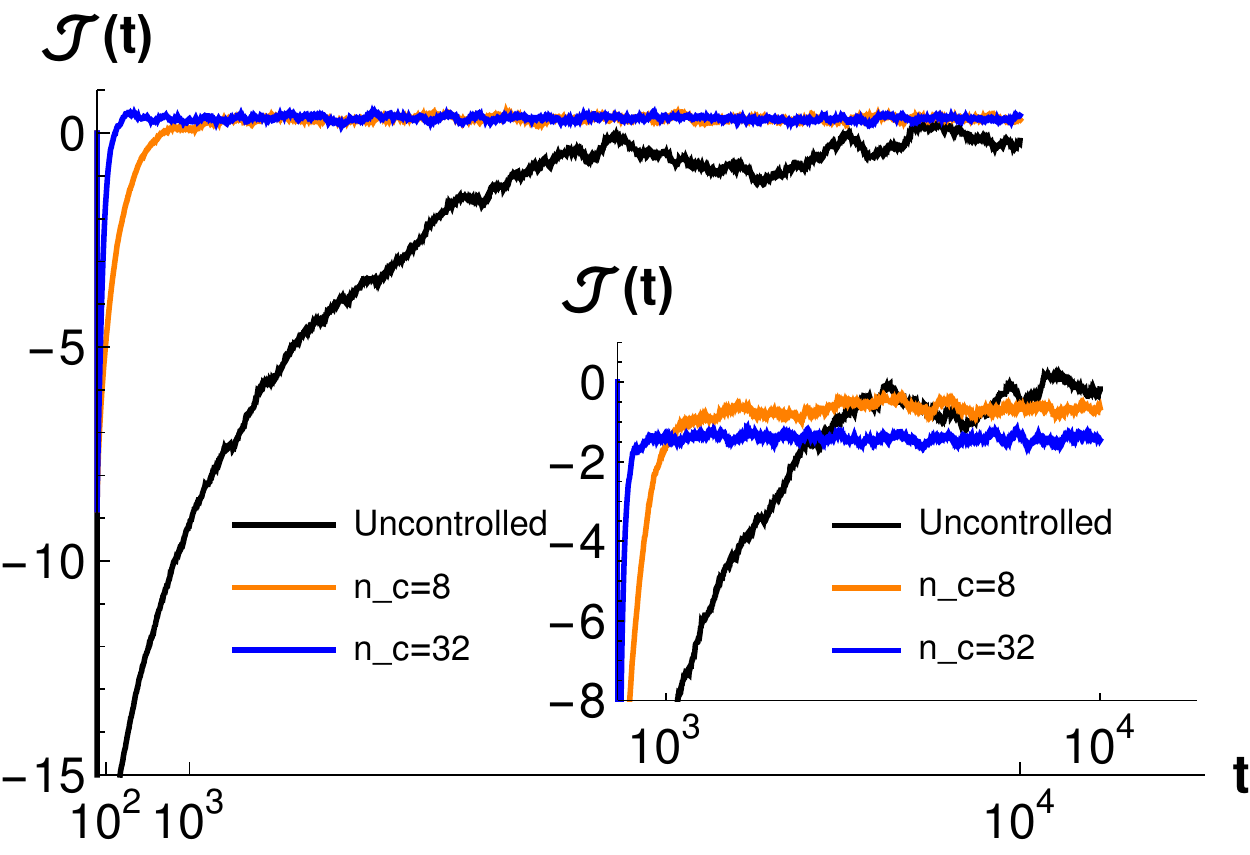}
% g1CostFunction.pdf: 0x0 px, 300dpi, 0.00x0.00 cm, bb=
\caption{The plot shows the time-dependent cost function \eref{J_eq} for the controlled KPZ growth process for $g=1$. 
The main panel shows the change in cost function ${\mathcal{J}}(t)$ for the saturation width $w_d=3$ and two different values of $n_c$. In the inset, the curve ${\mathcal{J}}(t)$ is plotted for $w_d=4$. The system parameters are $L=512$ and $\delta t = 0.0075$. The data are averaged over $3000$ different noise realizations.}
\label{J_fig}
\end{figure}
We also investigate a quantity that can be interpreted as a cost function ${\cal{J}}(t)$ for the roughness dynamics. We define it as the deviation of the square of the roughness from the desired saturation value of the roughness,
  \begin{equation}
  { \cal{J}}(t)=\sum_{x=1}^{L}[(h(x,t) - \langle h \rangle)^2-{w^2_d}]~,
  \label{J_eq}
  \end{equation}
where $w_d$ is the desired saturation width. For the uncontrolled KPZ growth, we insert $w^2_d=L/24$ to evaluate the cost function and observe that in the large-time limit, the function $ {\cal J}(t\rightarrow \infty)$ indeed approaches the value zero. 
%A zero variance indicates that the stationary state roughness is close to the mean roughness value of $L/24$. 
For the controlled cases,  our numerical simulations also confirm $ {\cal J}(t\rightarrow \infty)\approx0$ (see Fig. \ref{J_fig}) for all parameter ranges, when the desired saturation width is achieved, as is the case in the main figure with $w_d=3$, $g=1$, and $n_c=16,32$. 
The inset illustrates a situation, see also Fig. \ref{fig:g4_n_cwidth}, with $w_d=4$, $g=1$, and $n_c=32$, where the saturation width remains slightly below the desired value, yielding ${\cal{J}}<0$
for all times.
Hence, in controlling the saturation width for systems where $\lambda_c$ cannot be evaluated, we can use the cost function to optimize the growth process. The technique of optimizing the cost function to reach  ${\cal J}(t)\rightarrow0$ might provide a useful tool to control fluctuations in various complex systems.

\section{Conclusions}
\label{Conclusions}
This work constitutes progress towards a comprehensive understanding of the impact of controlling the surface roughness in stochastic growth processes. In this paper, we have presented a linear control scheme where one maintains the inherent properties of the controlled growth process, which is very different from previously discussed control schemes. Unlike the already known strategies, in this present scheme we have maintained the non-linearity of the underlying growth process while controlling the roughness to reach some desired value. Interestingly, due to the presence of the non-linearity in both controlled and uncontrolled Fourier modes, we can find some regimes where non-trivial scaling of the underlying growth process is maintained even under the controlled dynamics. 
Controlling the surface roughness through this scheme produces an intricate coupling via the non-linearity of the controlled and uncontrolled Fourier modes, which limits the accessible range of target values for the saturation roughness. We have employed a perturbative approach to obtain the steady-state controlled roughness. Interestingly, this explicit calculation provides us with information about the parametric regimes where the system can be controlled through linear control.

We have observed the controlled dynamics to be governed by the Edwards--Wilkinson universality class in a weak-coupling regime for a small number $n_c$ of controlled Fourier modes. Also for small $n_c$, in the case of large $g$, the dynamics displays KPZ universal behavior. Most importantly, the universal scaling here is identical with the corresponding intrinsic behavior. The only drawback of this scheme is that the upper saturation roughness is limited by the system size-dependent saturation width $\sqrt{L/24}$ for large $g$.
%; however, for small $g$ and $w_d>\sqrt{L/24}$, the dynamics at late times becomes ballistic.
Furthermore, we find that the probability distribution of the height function is positively skewed, and hence the depths of the valleys over the surface are more concentrated towards smaller values. 
%When $w_d<\sqrt{L/24}$, we find that the degree of asymmetry of the height distribution is small; thus, the controlled growth dynamics is similar to that of the uncontrolled system. 
We believe that this study that implements a linear control scheme provides a promising technique to control the stationary-state properties of a non-linear stochastic process while maintaining the dynamical scaling according to the universality class of the unperturbed system.

\appendix
 \section{Time-dependent width}
 \label{appendix}
 Here, we write down all existing terms upto ${\mathcal{O}}(g^2)$  in order to evaluate the surface roughness from eq.~(\ref{width0}). The zeroth-order term in $g$ of roughness is given as,
 \begin{align}
  \bar{\sigma}^0(L,s,s')&=\frac{1}{L^2}\sum_{k,k'}\frac{\hat{\eta}_c(k,s)\hat{\eta}_c( k',s)}{(s+\lambda)(s'+\lambda)} + \frac{1}{L^2}\sum_{ k,k'} \frac{\hat{\eta}(k,s)\hat{\eta}(k',s)}{(s+(2\pi k/L)^2)(s'+(2\pi k'/L)^2)}\no\\
  &=\frac{1}{L}\sum_{k,k'}\frac{\delta_{k,-k'}}{(s+s')(s+\lambda)(s'+\lambda)}\no\\
  &~~~+ \frac{1}{L}\sum_{k,k'} \frac{\delta_{k,-k'}}{(s+s')(s+(2\pi k/L)^2)(s'+(2\pi k'/L)^2)}~.
  \label{zero_ss}
 \end{align}
 Performing a Laplace back-transform, we can exactly solve the equal-time roughness which takes on the following form,
 \begin{equation}
 \sigma^0(L,t)=\frac{2}{L}\sum_{k=1}^{n_c}\frac{1-e^{-2~\lambda~t}}{2~\lambda}+\frac{2}{L}\sum_{k=n_c}^{L/2}\frac{1-e^{-2 ~(2\pi k/L)^2 ~t}}{2~(2\pi k/L)^2}~.
\end{equation}
Then to ${\mathcal{O}}(g^2)$, the controlled eq.~\eref{modes_c} gives 
\begin{align}
 \bar{\sigma}_{c}(L,s,s')&=\frac{1}{L^4}\sum_{m,m'}\frac{(g/2)^2}{(s+\lambda)(s'+\lambda)}\sum_{q,\gamma}\sum_{p,\gamma'}\left(\frac{2\pi}{L}\right)^4q(k+q)~p\cdot(k'-p)\no\\
&~~~\times\frac{\langle\hat{\eta}_c(q,\gamma)~\hat{\eta}_c(k-q,s-\gamma)~\hat{\eta}_c(p,\gamma')~\hat{\eta}_c(k'-p,s'-\gamma')\rangle}{(\gamma+\lambda)~[(s-\gamma)+\lambda]~(\gamma'+\lambda)~[(s'-\gamma')+\lambda]}\no
\\%  hchc
&~~~+\frac{1}{L^4}\sum_{k,k'}\frac{(g/2)^2}{(s+\lambda)(s'+\lambda)}\sum_{q,\gamma}\sum_{p,\gamma'}\left(\frac{2\pi}{L}\right)^4q(k+q)~p\cdot(k'-p)\no\\
&~~~~\times\frac{\langle\hat{\eta}_c(q,\gamma)~\hat{\eta}_u(k-q,s-\gamma)~\hat{\eta}_c(p,\gamma')~\hat{\eta}_u(k'-p,s'-\gamma')\rangle}{(\gamma+\lambda)~[(s-\gamma)+(2\pi/L)^2(k-q)^2]~(\gamma'+\lambda)~[(s'-\gamma')+(2\pi/L)^2(k'-p)^2]}\no
\\%hchu
&~~~+\frac{1}{L^4}\sum_{k,k'}\frac{(g/2)^2}{(s+\lambda)(s'+\lambda)}\sum_{q,\gamma}\sum_{p,\gamma'}\left(\frac{2\pi}{L}\right)^4q(k+q)~p\cdot(k'-p)\no\\
&~~~~ \times\frac{\langle\hat{\eta}_u(q,\gamma)~\hat{\eta}_c(k-q,s-\gamma)~\hat{\eta}_u(p,\gamma')~\hat{\eta}_c(k'-p,s'-\gamma')\rangle}{(\gamma+(2\pi q/L)^2)~[(s-\gamma)+\lambda]~(\gamma'+(2\pi p/L)^2)~[(s'-\gamma')+\lambda]}\no
\\% huhc
&~~~+\frac{1}{L^4}\sum_{k,k'}\frac{(g/2)^2}{(s+\lambda)(s'+\lambda)}\sum_{q,\gamma}\sum_{p,\gamma'}\left(\frac{2\pi}{L}\right)^4q(k+q)~p\cdot(k'-p)\no\\
&~~~~~\times\frac{\langle\hat{\eta}_u(q,\gamma)~\hat{\eta}_u(k-q,s-\gamma)\rangle~}{(\gamma+(2\pi q/L)^2)~[(s-\gamma)+(2\pi/L)^2(k-q)^2]}\no\\
&~~~~~\times\frac{\langle\hat{\eta}_u(p,\gamma')~\hat{\eta}_u(k'-p,s'-\gamma')\rangle}{(\gamma'+(2\pi p/L)^2)~[(s'-\gamma')+(2\pi/L)^2(k'-p)^2]}~.
\label{g2_cc}
\end{align}
In the above expression, summations in reciprocal space only range from $1$ to $n_c$. 

Similarly we obtain the following four terms for the uncontrolled part given by eq.~\eref{modes_u}, 
\begin{align}
 \bar{\sigma}_{u}(L,s,s')&=\frac{1}{L^4}\sum_{k,k'}\frac{(g/2)^2}{(s+(2\pi k/L)^2)(s'+(2\pi k'/L)^2)}\sum_{q,\gamma}\sum_{p,\gamma'}\left(\frac{2\pi}{L}\right)^4q(k+q)~p\cdot(k'-p)\no\\
&~~~\times\frac{\langle\hat{\eta}_c(q,\gamma)~\hat{\eta}_c(k-q,s-\gamma)~\hat{\eta}_c(p,\gamma')~\hat{\eta}_c(k'-p,s'-\gamma')\rangle}{(\gamma+\lambda)~[(s-\gamma)+\lambda]~(\gamma'+\lambda)~[(s'-\gamma')+\lambda]}\no
\\%  hchc
&~~~+\frac{1}{L^4}\sum_{k,k'}\frac{(g/2)^2}{(s+(2\pi k/L)^2)(s'+(2\pi k'/L)^2)}\sum_{q,\gamma}\sum_{p,\gamma'}\left(\frac{2\pi}{L}\right)^4q(k+q)~p\cdot(k'-p)\no\\
&~~~~\times \frac{\langle\hat{\eta}_c(q,\gamma)~\hat{\eta}_u(k-q,s-\gamma)~\hat{\eta}_c(p,\gamma')~\hat{\eta}_u(k'-p,s'-\gamma')\rangle}{(\gamma+\lambda)~[(s-\gamma)+(2\pi/L)^2(k-q)^2]~(\gamma'+\lambda)~[(s'-\gamma')+(2\pi/L)^2(k'-p)^2]}\no
\\%hchu
&~~~+\frac{1}{L^4}\sum_{k,k'}\frac{(g/2)^2}{(s+(2\pi k/L)^2)(s'+(2\pi k'/L)^2)}\sum_{q,\gamma}\sum_{p,\gamma'}\left(\frac{2\pi}{L}\right)^4q(k+q)~p\cdot(k'-p)\no\\
&~~~~\times \frac{\langle\hat{\eta}_u(q,\gamma)~\hat{\eta}_c(k-q,s-\gamma)~\hat{\eta}_u(p,\gamma')~\hat{\eta}_c(k'-p,s'-\gamma')\rangle}{(\gamma+(2\pi q/L)^2)~[(s-\gamma)+\lambda]~(\gamma'+(2\pi p/L)^2)~[(s'-\gamma')+\lambda]}\no\\% huhc
&~~~+\frac{1}{L^4}\sum_{k,k'}\frac{(g/2)^2}{(s+(2\pi k/L)^2)(s'+(2\pi k'/L)^2)}\sum_{q,\gamma}\sum_{p,\gamma'}\left(\frac{2\pi}{L}\right)^4q(k+q)~p\cdot(k'-p)\no\\
&~~~~~\times \frac{\langle\hat{\eta}_u(q,\gamma)~\hat{\eta}_u(k-q,s-\gamma)\rangle}{(\gamma+(2\pi q/L)^2)~[(s-\gamma)+(2\pi/L)^2(k-q)^2]}\no\\
&~~~~~\times\frac{\langle\hat{\eta}_u(p,\gamma')~\hat{\eta}_u(k'-p,s'-\gamma')\rangle}{(\gamma'+(2\pi p/L)^2)~[(s'-\gamma')+(2\pi/L)^2(k'-p)^2]}~.
\label{g2_UU}
\end{align}
Performing the contraction over the noise, the Laplace-transformed roughness to second order in $g$ takes on the following expression with $2 n_c$ controlled Fourier modes:
\begin{eqnarray}
\hspace{-2cm} \bar{\sigma}_{c}(L,s,s')&=&\frac{1}{L^4}\sum_{k,k'}\frac{(g/2)^2}{(s+\lambda)(s'+\lambda)}\sum_{q,\gamma}\sum_{p,\gamma'}\left(\frac{2\pi}{L}\right)^4q(k+q)~p\cdot(k'-p)\no\\
&&~\times \frac{\delta_{q,-p}\delta_{k-p,-k'+q}}{(\gamma+\lambda)~[(s-\gamma)+\lambda]~(\gamma'+\lambda)~[(s'-\gamma')+\lambda](\gamma+\gamma')}\no\\%  hchc
&&~\times \frac{1}{(s+s'-(\gamma+\gamma'))}+\frac{1}{L^4}\sum_{k,k'}\frac{(g/2)^2}{(s+\lambda)(s'+\lambda)}\no\\
&&~\times \sum_{q,\gamma}\sum_{p,\gamma'}\frac{q(k+q)~p\cdot(k'-p)~\delta_{q,-p}\delta_{k-p,-k'+q}}{(\gamma+\lambda)~[(s-\gamma)+(k-q)^2]~(\gamma'+\lambda)}\no\\
&&~\times \frac{1}{~[(s'-\gamma')+(k'-p)^2](\gamma+\gamma')(s+s'-(\gamma+\gamma'))}\no\\%hchu
&+&\frac{1}{L^4}\sum_{k,k'}\frac{(g/2)^2}{(s+\lambda)(s'+\lambda)}\sum_{q,\gamma}\sum_{p,\gamma'}\left(\frac{2\pi}{L}\right)^4q(k+q)~p\cdot(k'-p)\no\\
&&~\times \frac{\delta_{q,-p}\delta_{k-p,-k'+q}}{(\gamma+q^2)~[(s-\gamma)+\lambda]~(\gamma'+p^2)~[(s'-\gamma')+\lambda](\gamma+\gamma')}\no\\% huhc
&&~\times\frac{1}{(s+s'-(\gamma+\gamma'))}+\frac{1}{L^4}\sum_{k,k'}\frac{(g/2)^2}{(s+\lambda)(s'+\lambda)}\no\\
&&~\times\sum_{q,\gamma}\sum_{p,\gamma'}\frac{q(k+q)~p\cdot(k'-p)~\delta_{q,-p}\delta_{k-p,-k'+q}}{(\gamma+q^2)~[(s-\gamma)+(k-q)^2]~(\gamma'+p^2)}\no\\
&&~\times \frac{1}{~[(s'-\gamma')+(k'-p)^2](\gamma+\gamma')(s+s'-(\gamma+\gamma')))}~.
\label{g2_CC2}
\end{eqnarray}

Similarly, we replace the noise with a delta function in eq.~\eref{g2_UU} pertaining to the terms for the uncontrolled part of the coupled equations~\eref{modes_u}. The Laplace transforms of all the terms above yield an exponential solution. Nevertheless, the time-dependent solution of the ${\mathcal O}(g^2)$ terms in the $t\rightarrow0$ limit is proportional to $ t^4$ which conveys that once the control sets in, the dynamic growth becomes very fast. While we do see this faster growth kinetics in our numerics, we did not readily observe a $t^4$ behavior because of the small prefactor of the solution which delays the onset of the $t^4$ behavior depending on $n_c$. Our calculation suggests that the dominant contribution in the controlled stationary-state surface roughness originates from the term $h_c(k,t)h_c(k',t)$, hence the time-dependent roughness up to ${\mathcal O}(g^2)$ can be approximated as
\begin{align}
&\sigma (L,t) = {\sigma}^0(L,t)+ {\sigma}_c(L,t) %\\ &&
=\frac{2}{L}\sum_{k=1}^{n_c}\frac{1-e^{-2~\lambda~t}}{2~\lambda}+\frac{2}{L}\sum_{k=n_c}^{L/2}\frac{1-e^{-2 ~(2\pi k/L)^2 ~t}}{2~(2\pi k/L)^2}\no\\
&-\frac{g^2 (2\pi)^2}{L^4}\sum_{k=1}^{n_c}\sum_{p=1}^{nc}\frac{ (2\pi/L)^4p^2 (k - p)^2}{ 3 \lambda^4}~e^{-4\lambda t}%\nonumber\\ &&~~\times 
\Bigl[ 3-16~e^{\lambda t}+e^{4\lambda t}-12~e^{2\lambda t}(t \lambda-1) \Bigr] . 
\end{align}
In the infinite-time limit this expression becomes identical to eq.~\eref{second-order_sol} for the desired squared surface roughness.

\ack 
Research was sponsored by the Army Research Office and was accomplished under Grant Number \textbf{W911NF-17-1-0156}. 
The views and conclusions contained in this document are those of the authors and should not be interpreted as representing the official policies, either expressed or implied, of the Army Research Office or the U.S. Government. The U.S. Government is authorized to reproduce and distribute reprints for Government purposes notwithstanding any copyright notation herein.

\section*{References}
\providecommand{\newblock}{}

%\bibliography{roughness}
%\bibliographystyle{iopart-num}
\end{document}